# Spontaneous Symmetry Breaking and Emergent Helicity in Achiral $D_{4h}$-Symmetric Zinc Phthalocyanine Condensates

Bruno S. Zanatta,[1,*] Giovani B. de Oliveira,[1] Marcos E. G. Carmo,[2] Antonio Eduardo H. Machado,[2,3] Thiago F. Silva,[3] Guilherme F. S. Miguel,[4] Antonio Otavio T. Patrocinio,[2] Raigna A. da Silva,[1] Erick Piovesan,[1] and Alexandre Marletta[1,†]

[1]Physics Institute, Federal University of Uberlandia, 38400-902, Uberlandia, MG, Brazil
[2]Chemistry Institute, Federal University of Uberlandia, 38400-902, Uberlandia, MG, Brazil
[3]Physics Institute, Federal University of Catalão, 75704-020 Catalão, GO, Brazil
[4]Department of Electronic Computation, Federal University of Espirito Santo, 29932-540 São Mateus, ES, Brazil

Herein it is shown spontaneous symmetry breaking in supramolecular aggregates of $D_{4h}$-symmetric zinc phthalocyanine (ZnPc). *Ab initio* density functional theory calculations at the M06/DGDZVP level reveal that intermolecular interactions induce a subtle relaxation of the macrocyclic framework, producing a characteristic red shift of the aza-bridge (C–N–C) stretching mode and symmetric stretching of the pyrrole ring. Confocal Raman Optical Activity measurements further reveal a negative Cotton effect at 1505 $cm^{-1}$ and 1338 $cm^{-1}$, providing evidence of emergent supramolecular chirality. Our findings identify pristine ZnPc as a model system for spontaneous self-assembled chiral symmetry breaking in molecular condensates and suggest new opportunities for chiroptical, spin-selective, and quantum functional materials.

## I. INTRODUCTION.

Symmetry breaking in topological and supramolecular organic semiconductor materials plays a fundamental role in low-dimensional correlation processes, providing selectivity to several spin-polarized transport phenomena [1–3]. The emergence of homochiral supramolecular architectures assembled from topological achiral building blocks highlights the fundamental role of intermolecular interactions in driving spontaneous symmetry breaking [2,3]. While biological macromolecules achieve such polarization through pre-existing stereocenters, generating macroscopic chiral continua within pristine, unfunctionalized molecular aggregates remains a strategic frontier in metamaterials and optical spintronics [1,3,4]. In dense molecular condensates (molecular aggregates), this symmetry reduction is mediated exclusively by weak, non-covalent interactions [2,5,6]. Recently, enantiomorphism can manifest not only via static structural conformations, but also through purely electronic mechanisms, such as orbital-specific chiral distortions induced by localized asymmetric charge transfer or by asymmetric molecular arrangements [4,6,7]. A molecular model system for investigating this fundamental interaction is zinc phthalocyanine (ZnPc), a highly symmetric macrocyclic architecture with a square-planar monomeric form belonging to the $D_{4h}$ point group [7–9]. While individual gas-phase monomers are inherently achiral, phthalocyanine derivatives can overcome point-group symmetry constraints upon aggregation, self-organizing into chiral supramolecular architectures governed by an intricate balance of hydrogen bonding, van der Waals interactions, and π-π stacking [3,5,6,10]. Recent structural studies on macrocyclic Langmuir-Blodgett films have demonstrated the emergence of supramolecular chirality upon film formation [11].

In this Letter, we show that intrinsically achiral zinc phthalocyanine (ZnPc) molecules self-assemble into a chiral supramolecular architecture without chemical modification or external confinement. The ZnPc molecule was characterized using absorption (Abs), photoluminescence (PL), photoluminescence excitation (PLE), and Raman spectroscopy to validate the structure of the material. *Ab initio* density functional theory (DFT) calculations predict a red shift of the aza-bridge (C–N–C) stretching mode and symmetric stretching of the pyrrole ring upon aggregation, which is confirmed by optical and vibrational spectroscopies. Emission ellipsometry (EE) probes the emergence of macroscopic chirality in the intrinsically achiral ZnPc system. Confocal Raman Optical Activity (Confocal-ROA) measurements reveal a negative Cotton effect at 1505 $cm^{-1}$ and 1338 $cm^{-1}$, providing direct evidence of spontaneous symmetry breaking and emergent supramolecular chirality in pristine ZnPc aggregates.

## II. METHODS

Commercial zinc phthalocyanine (ZnPc, Sigma-Aldrich $M_W$ = 577.92 g/mol, CAS 14320-04-8, Lot & Filling code: 1187020 53605201) was used without further purification. Solutions with concentrations $10^{-6}$ (C6), $10^{-5}$ (C5), and $10^{-4}$ mol $L^{-1}$ (C4), together with a saturated solution (C3) and powder sample, were prepared in dimethyl sulfoxide (DMSO, P.A.-A.C.S., ≥ 99.9%, $M_W$ = 78.13 g/mol, Synth, Code: D1011.01.BJ, Lot: 200239) (Fig. S1 in Ref. [12]). Additional characterization data, including elemental

*Contact author: brunozanatta@ufu.br
†Contact author: marletta@ufu.br

analysis and ICP-OES measurements, confirm sample purity [12].

UV-Vis spectra were recorded using a FEMTO 800XI UV-VIS spectrophotometer within a spectral range of 190 to 1100 nm. PL and PLE measurements were performed using a Shimadzu RF-5301pc Spectro fluorophotometer. Selective excitation was performed using a diode laser as an excitation source in 635 nm and an Ocean Optics USB4000 spectrophotometer. EE measurements [Fig S4 in Ref. [12]] were performed using Fourier series analysis to calculate the Stokes Parameters [12] (see also ref. [13] therein). Raman scattering measurements were carried out using a Horiba Jobin Yvon Olympus BX41 microscope coupled with a uEye camera, a Horiba Jobin Yvon iHR550, a Synapse CCD detector, and a 514.5 nm Coherent Innova 24C excitation laser. Confocal-ROA spectra were collected using the same setup supplemented with an achromatic linear polarizer and a quarter-wave plate [12] (see also Ref. [14] therein).

The PL spectra were fitted using the Luminescence Spectrum Line Shape (LSLS) software [15]. The lineshape analysis considers the experimental data zero-phonon transition energy $E_{00}$, phonon energies $E_{01}$, $E_{02}$, and $E_{03}$, homogeneous linewidth $d$, and adjustable Huang-Rhys parameters $S_1$, $S_2$, and $S_3$ [12].

*Ab initio* DFT calculations were performed using the Gaussian 09 software package (Revision E.01) [16]. Geometry optimizations and vibrational frequency calculations, including Raman intensities, were carried out using the M06 functional in conjunction with the DGDZVP basis set [17]. The initial geometry of the tetrameric system was constructed from the previously optimized monomeric structure at the same level of theory, using an initial intermolecular separation of approximately 4 Å between adjacent monomer units.

## III. RESULTS

UV-Vis absorption spectra of the C6 sample showed excellent agreement with those reported in the literature [18]. A red-shift of the PL emission maximum (675–700 nm) and intensity enhancement of the lower-energy band (725–775 nm) were observed as a function of the sample concentration, as shown in Fig 1. These results confirm the progressive formation of aggregated states. PLE spectra corroborate with this behavior, clearly defining the transition from isolated monomers at low concentrations (C6) to H-type aggregates in the saturated regime (C3), as evidenced by the blue-shift of the Q-band [Fig. 1].

The LSLS analysis reveals that increasing ZnPc concentration promotes a progressive red shift of the $E_{00}$ transition and a concomitant reduction of the disorder parameter ($d$), evidencing the stabilization and structural ordering of the aggregated state [12]. The vibronic energies $E_{01}$ and $E_{02}$ exhibit pronounced mode switching, indicating that aggregation selectively suppresses vibrational modes associated with isolated molecules while activating alternative vibronic channels [12]. In contrast, $E_{03}$ remains optically active throughout the aggregation process, displaying only a moderate red shift (~25 cm$^{-1}$), which suggests that this high-frequency vibration is perturbed, but not quenched, by intermolecular π-π interactions [12]. Collectively, these results demonstrate that supramolecular assembly significantly modifies the vibronic coupling and governs the radiative relaxation pathways in ZnPc aggregates. Figure 1 presents the absorption (C5), PL and PLE (C3 and C6) spectra for ZnPc samples.

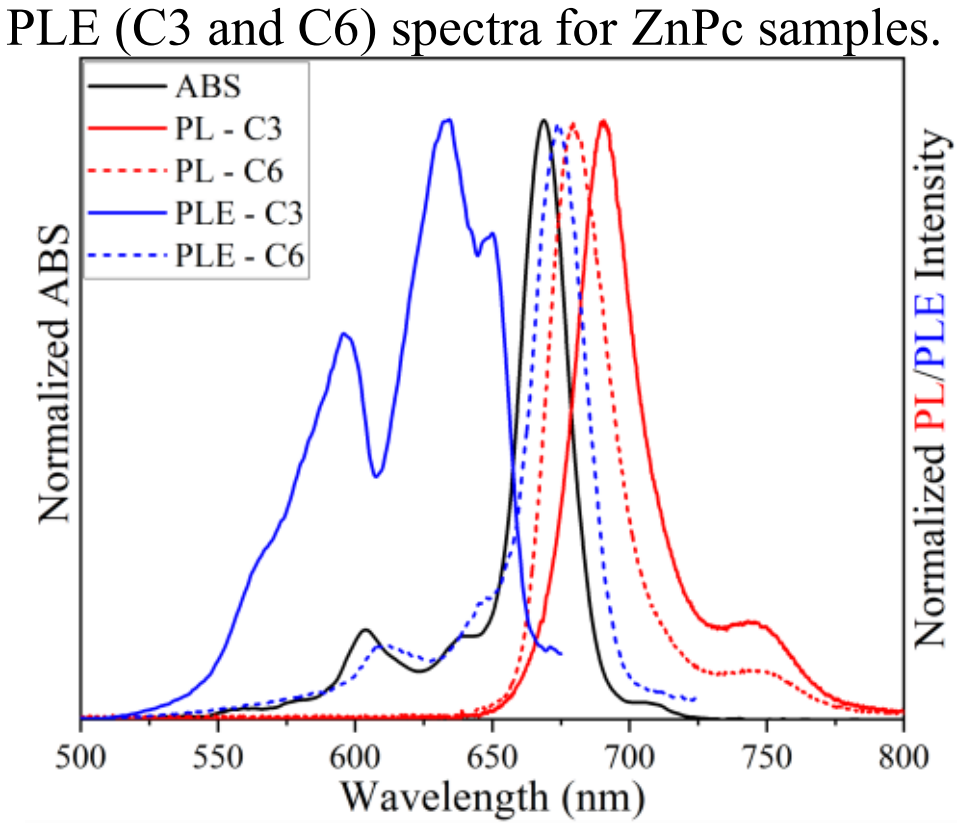

FIG. 1. (Left) Normalized absorption spectra of ZnPc in a DMSO solution at concentration C5. (Right) Normalized PL and PLE spectra of samples C6 and C3.

The two-species LSLS model further quantifies the relative contributions of isolated monomers and aggregated states through the statistical weights $K_1$ and $K_2$, respectively [12]. As the ZnPc concentration increases, $K_1$ progressively decreases while $K_2$ increases, reflecting the gradual conversion from monomeric to aggregated species [12]. Although the actual system may comprise a continuum of intermediate configurations, the two-species approximation captures the dominant structural extremes and reproduces the observed evolution of the vibronic coupling. These results corroborate the suppression and switching of vibrational modes discussed above and demonstrate that aggregate formation progressively dictates the dominant radiative relaxation channel.

Raman spectroscopy was employed to experimentally probe the structural modifications induced by aggregation. The Raman spectrum of the pristine ZnPc sample agrees well with literature assignments [Fig. S7 in Ref. [12]], validating the mode identification. Both the saturated solution (C3) and the powder sample exhibit a red shift of the 1505 cm$^{-1}$ aza-bridge stretching mode and the 1338 cm$^{-1}$ symmetric stretching of the pyrrole ring, as shown in Fig. 2. This frequency shift indicates a relaxation of the C–N–C bridges driven by intermolecular $\pi$-$\pi$ interactions within the aggregate.

*Contact author: brunozanatta@ufu.br
†Contact author: marletta@ufu.br

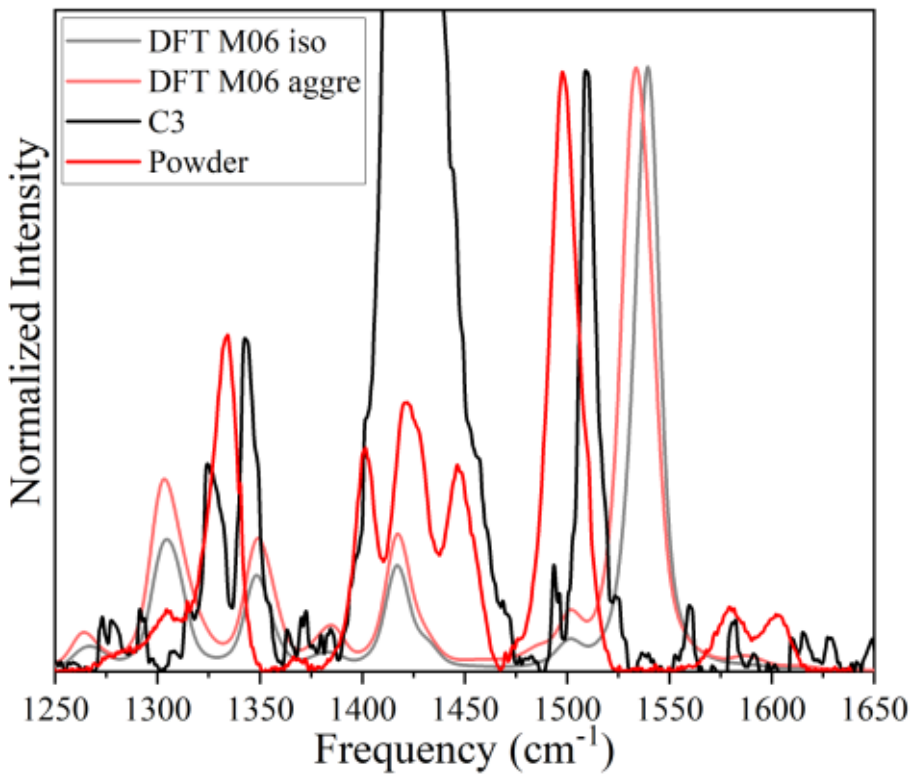

FIG. 2. Experimental Raman spectra of bulk ZnPc powder and its DMSO-diluted solution, plotted alongside DFT-calculated theoretical spectra. The comparison highlights the coordinate frequency shift of the 1505 cm$^{-1}$ aza-bridge (C–N–C) stretching mode and the 1338 cm$^{-1}$ symmetric stretching of the pyrrole ring (involving internal C–N and C–C bonds) induced by molecular aggregation.

DFT calculations were performed to investigate the microscopic origin of the emergent supramolecular chirality [Figs. S8–S10 and Table S5 in Ref. [12]]. The optimized structures of both the isolated ZnPc molecule and the aggregated model exhibited no imaginary frequencies, confirming that they correspond to stable local minima. The optimized geometrical parameters, vibrational frequencies, and Raman activities showed excellent agreement with experimental data [Table S5 in Ref. [12]]. The M06/DGDZVP approach successfully reproduced the experimentally observed red-shift of the aza-bridge vibrational mode, highlighting the importance of accurately describing intermolecular interactions in the aggregated state, as shown in Fig. 2. The excellent agreement between theory and experiment provides strong evidence that aggregation perturbs the macrocyclic framework and supports the spontaneous symmetry breaking inferred from the EE measurements.

The supramolecular chirality of the ZnPc assembly was established by Confocal-ROA. Unlike achiral systems, which do not exhibit a ROA response, chiral structures generate characteristic vibrational signatures associated with their handedness. Figure 3 presents the ROA and Raman spectra of the ZnPc powder. The observed ROA response constitutes direct evidence of spontaneous chiral symmetry breaking in the aggregated state. EE measurements, inserted in Fig. 3, were subsequently employed to probe the emergence of macroscopic chirality in the intrinsically achiral ZnPc system [12]. The extracted asymmetry parameter ($g$) remains close to zero for samples C6 and C5, consistent with an isotropic distribution of isolated molecules [Table S1 in Ref. [12]]. As displayed in the inset of Fig. 3, at higher concentrations (C3), $g$ becomes significant, indicating preferential orientational ordering and providing direct evidence of macroscopic symmetry breaking in the condensed phase.

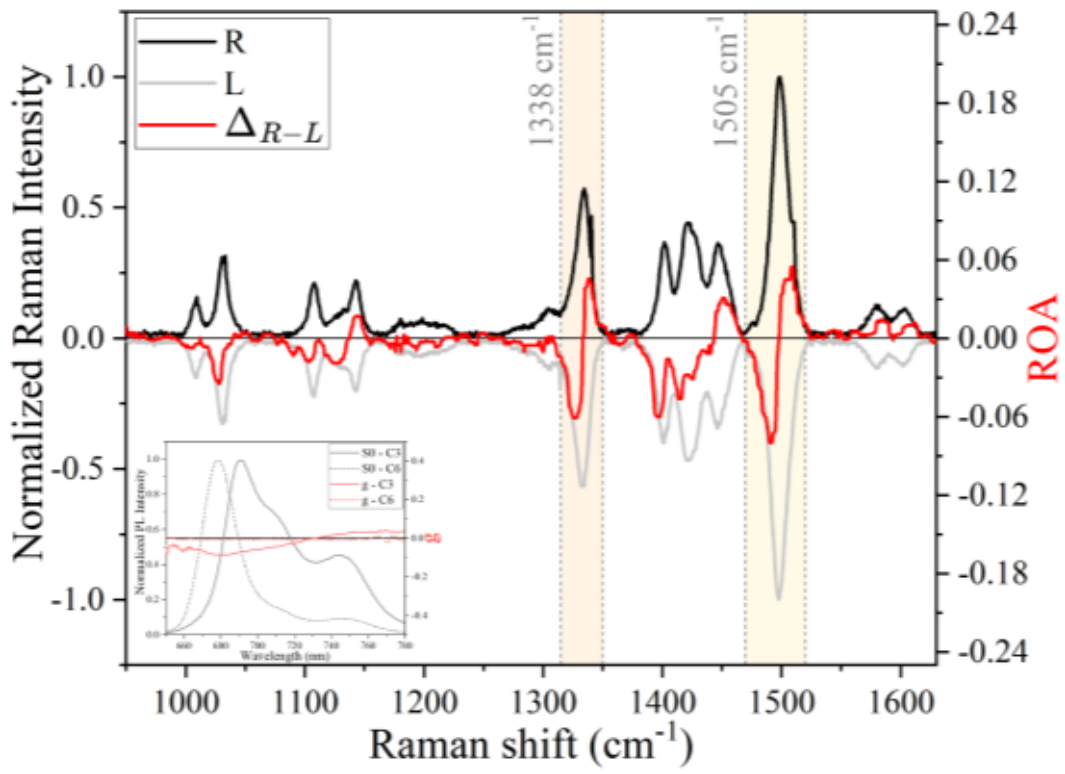

FIG. 3. Experimental Raman scattering (black/gray lines) and Confocal-ROA (red line) spectra of bulk ZnPc powder. Shaded regions highlight the vibrational modes exhibiting a pronounced Cotton effect in the ROA response. In the inset, EE spectra of S0 (black) and g (red) parameters for samples C6 (dash-line) and C3 (line).

As depicted in Figure 3, the aza-bridge stretching mode near 1505 cm$^{-1}$ and the symmetric stretching of the pyrrole ring at 1338 cm$^{-1}$ exhibit a distinct Cotton effect in the ROA spectrum of the bulk powder sample (red line). The analysis of ROA signs yields critical stereochemical information; specifically, the sequence of the bands dictates the structural handedness. Observing the obtained ROA curve (red line), a negative inflection point is evident, meaning a negative Cotton effect for the ZnPc aggregate. This spectral signature strongly indicates that the ZnPc molecules self-assemble into an inverted, left-handed (anticlockwise) supramolecular helix. The 3D model generated by Gaussian 09 is presented on the left-side in Fig S11 in Ref [12], while a schematic projection illustrating the helix formation from an alternative perspective is shown on the right. As previously discussed in the DFT analysis, the generated 3D model indicates that the aggregated state of ZnPc adopts a structural arrangement that deviates from conventional packing paradigms [11]. The observed negative Cotton effect provides direct spectroscopic evidence of spontaneous symmetry breaking of ZnPc. Together, the ROA measurements and theoretical 3D model demonstrate that intrinsically achiral zinc phthalocyanine molecules self-assemble into a chiral supramolecular architecture upon aggregation.

*Contact author: brunozanatta@ufu.br
†Contact author: marletta@ufu.br

## V. CONCLUSIONS

In summary, we demonstrate that intrinsically achiral zinc phthalocyanine (ZnPc) molecules spontaneously self-assemble into a chiral supramolecular architecture upon aggregation. DFT calculations predict a structural relaxation of the aza-bridge network, manifested as a red shift of the 1505 $cm^{-1}$ vibrational mode, which is directly confirmed by Raman spectroscopy. The emergence of supramolecular chirality is independently verified by Emission Ellipsometry and Confocal Raman Optical Activity measurements, revealing macroscopic symmetry breaking and a well-defined chiral handedness in the aggregated state. Furthermore, lineshape analysis of the photoluminescence spectra demonstrates that aggregation reorganizes the vibronic coupling and modifies the dominant radiative relaxation pathways. These findings establish pristine ZnPc as a model system for investigating spontaneous symmetry breaking in molecular condensates and open new opportunities for the design of chiroptical, spin-selective, and quantum functional materials.

## DATA AVAILABILITY

The data that support the findings of this study are available from the corresponding author upon reasonable request.

## ACKNOWLEDGMENTS

The authors gratefully acknowledge the financial support of the Brazilian research funding agency CNPQ (PQ 309984/2022-0, INCT-CIMOL 406804/2022-2), Minas Gerais Research Foundation (FAPEMG; Process Number 31280* RED 00081-23) and MCTI/FINEP/FNDCT (Grant No. 0966/24 #01.25.0086.00).

*Contact author: brunozanatta@ufu.br
†Contact author: marletta@ufu.br

*Contact author: brunozanatta@ufu.br
†Contact author: marletta@ufu.br

# Supplemental Material: Spontaneous Symmetry Breaking and Emergent Helicity in Achiral $D_{4h}$-Symmetric Zinc Phthalocyanine Condensates

Bruno S. Zanatta,[1,*] Giovani B. de Oliveira,[1] Marcos E. G. Carmo,[2] Antonio Eduardo H. Machado,[2,3] Thiago F. Silva,[3] Guilherme F. S. Miguel,[4] Antonio Otavio T. Patrocinio,[2] Raigna A. da Silva,[1] Erick Piovesan,[1] and Alexandre Marletta[1,†]

[1]Physics Institute, Federal University of Uberlandia, 38400-902, Uberlandia, MG, Brazil

[2]Chemistry Institute, Federal University of Uberlandia, 38400-902, Uberlandia, MG, Brazil

[3]Physics Institute, Federal University of Catalão, 75704-020 Catalão, GO, Brazil

[4]Department of Electronic Computation, Federal University of Espirito Santo, 29932-540 São Mateus, ES, Brazil

## I. Zinc Phthalocyanine samples

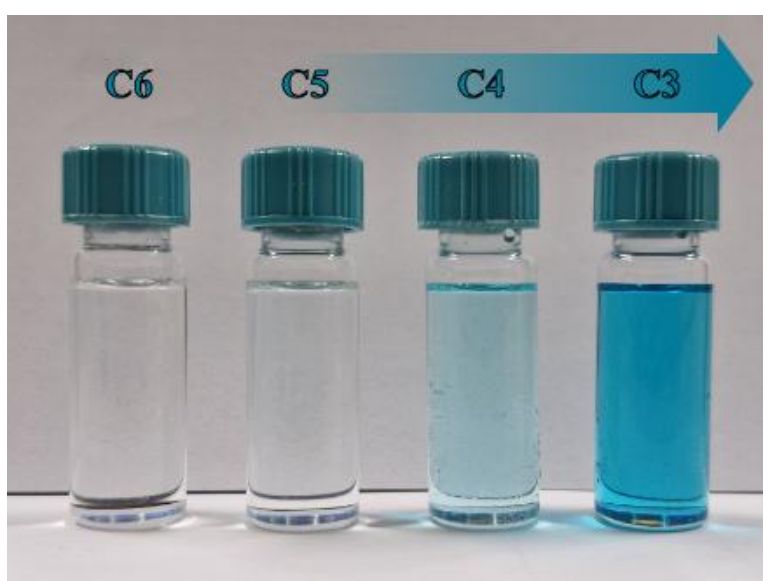

FIG. S1. ZnPc samples in DMSO solutions: C6 ($10^{-6}$ mol/L), C5 ($10^{-5}$ mol/L), C4 ($10^{-4}$ mol/L), and C3 saturated solutions.

## I. Chemical composition validation of Zinc Phthalocyanine.

Elemental analysis calculated for $C_{32}H_{16}N_8Zn$: C, 66.50%; H, 2.79%; N, 19.39%; found: C, 66.42%; H, 2.85%; N, 19.31%. The zinc concentration determined by ICP-OES was 11.25% (theoretical: 11.31%).

## IV. Photoluminescence and Photoluminescence Excitation spectra

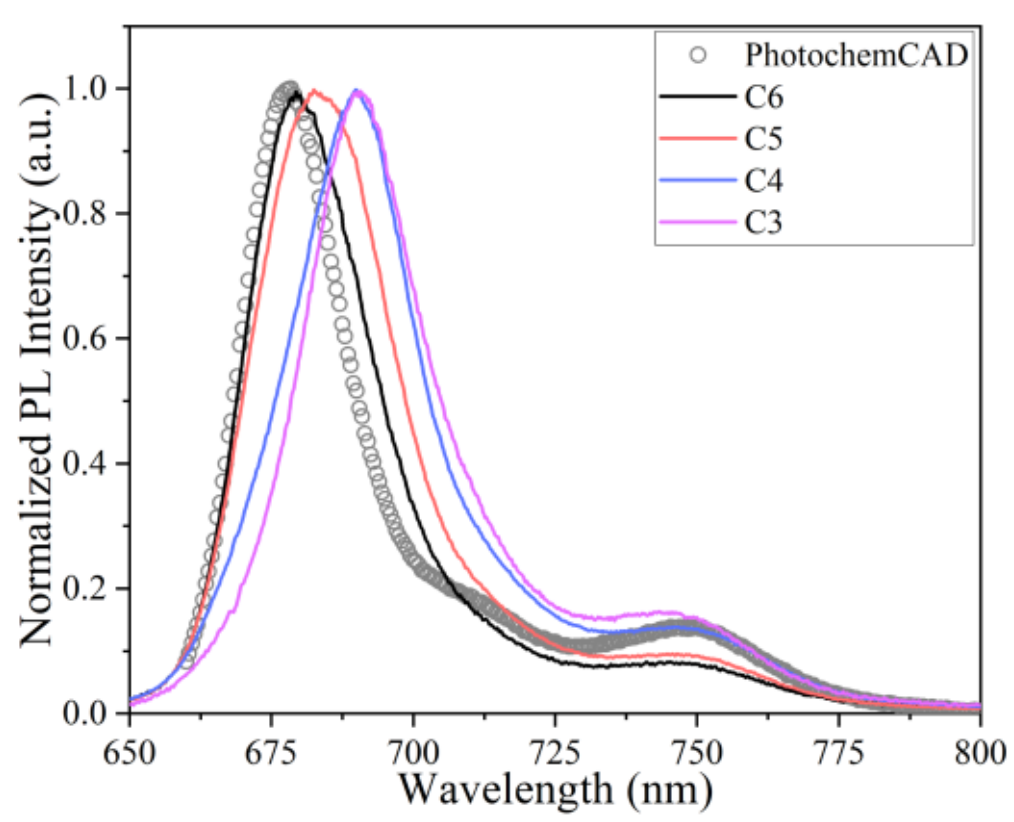

FIG. S2. Normalized Photoluminescent spectra for ZnPC samples diluted in DMSO in different concentrations, excited with 635 nm wavelengths.

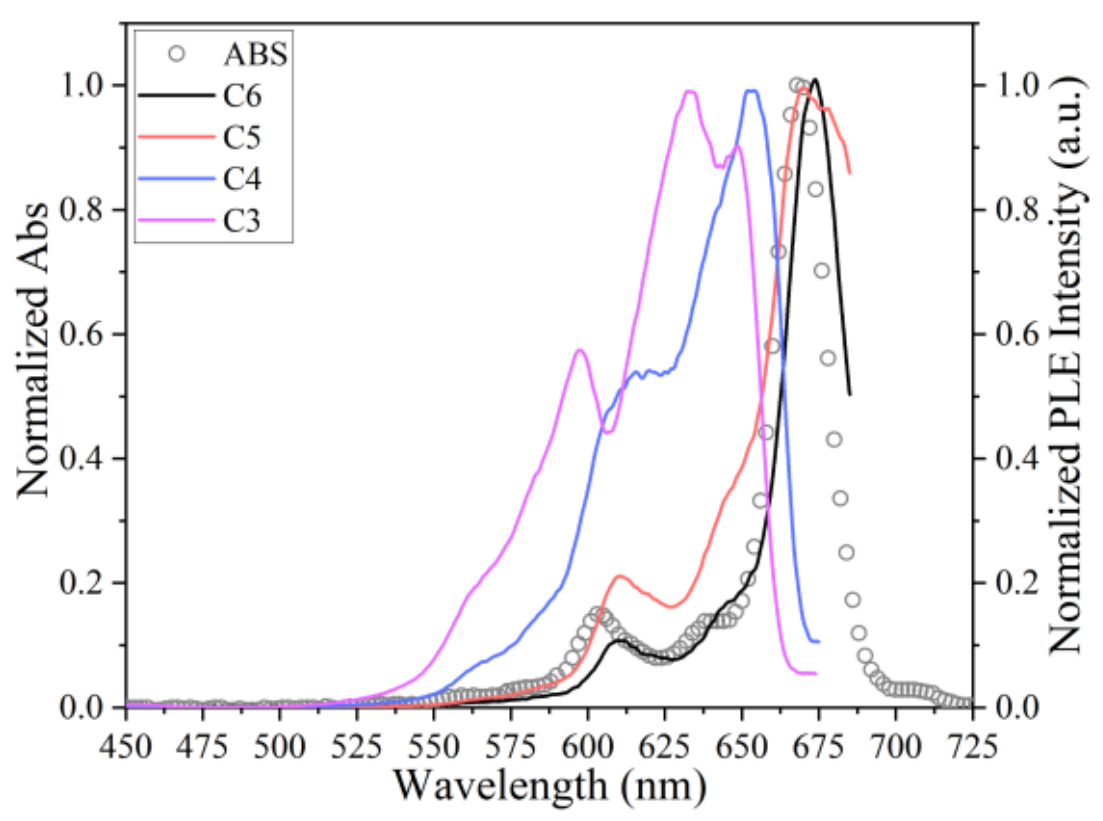

FIG. S3. Normalized Absorption and Photoluminescence Excitation of the ZnPC samples diluted in DMSO in different concentrations.

## II. Emission ellipsometry, experimental setup and equations.

The light intensity in an emission ellipsometry experiment is measured following the equation [13],

$$I(\theta) = \frac{1}{2}[S_0 + S_1 cos^2(2\theta) + S_2 cos(2\theta)sin(2\theta) - S_3 sin(2\theta)], \quad (1)$$

and the Stokes parameters $S_i$ $(i = 0,1,2,3)$ are [13]

$$S_0 = \frac{2}{N}\sum_{n=0}^{N} I(n\theta_j)[1 - 2cos(4n\theta_j)], \quad (2)$$

$$S_1 = \frac{8}{N}\sum_{n=0}^{N} I(n\theta_j)cos(4n\theta_j), \quad (3)$$

$$S_2 = \frac{8}{N}\sum_{n=0}^{N} I(n\theta_j)sin(4n\theta_j), \quad (4)$$

$$S_3 = \frac{4}{N}\sum_{n=0}^{N} I(n\theta_j)sin(2n\theta_j), \quad (5)$$

*Contact author: brunozanatta@ufu.br

†Contact author: marletta@ufu.br

with $N = 9$ and $\theta_j = 40°$. The degree of polarization in terms of the Stokes parameters is defined as [13]

$$p = \frac{(S_1{}^2+S_2{}^2+S_3{}^2)^{\frac{1}{2}}}{S_0}. \qquad (6)$$

The anisotropy factor $(r)$ and the asymmetry factor $(g)$ is also defined as [13]

$$r = \frac{-2\frac{S_1}{S_0}}{3+\frac{S_1}{S_0}}, \qquad (7)$$

$$g = 2\frac{S_3}{S_0}. \qquad (8)$$

The experimental setup was built according to the scheme in Fig. S4, and the data obtained from the experiment are presented in Table S1.

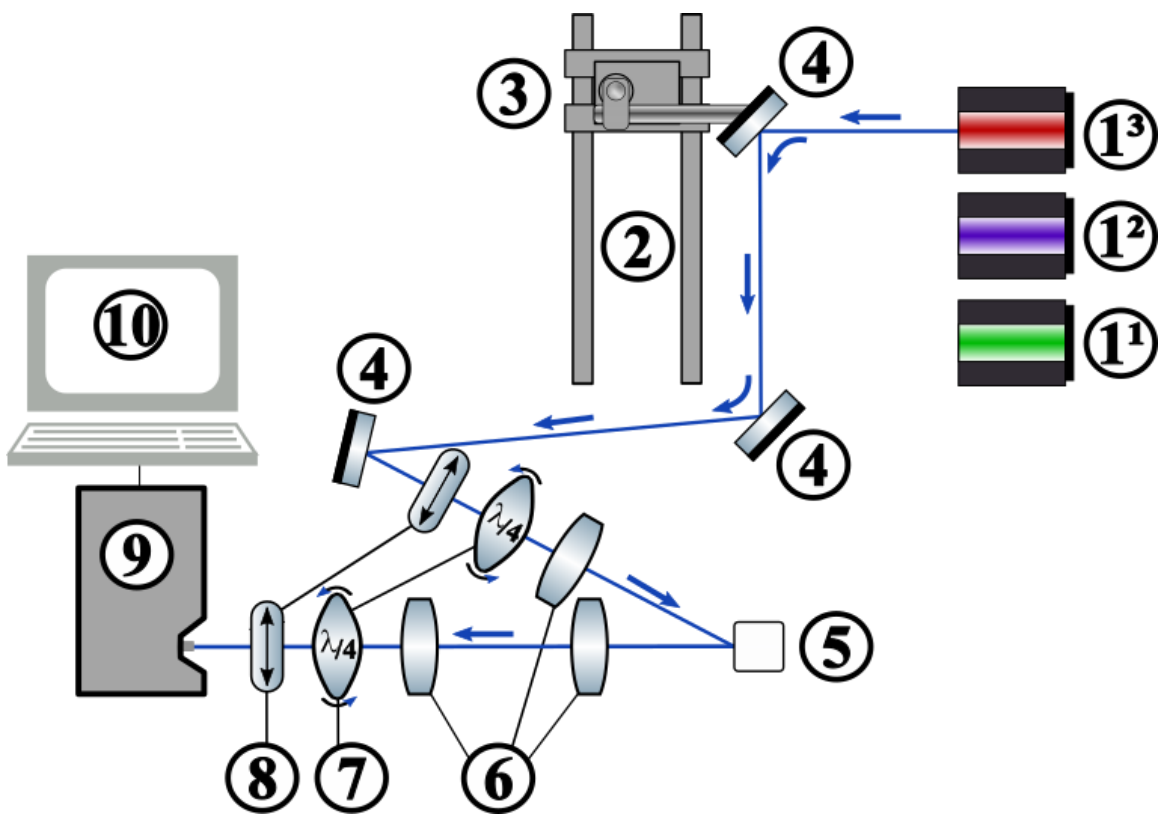

FIG. S4. Scheme to represent the Emission Ellipsometry setup. The numbers help identify the equipment and components used in the setup. 1ⁱ (i = 1, 2, 3) is the laser sources with (1) 514 nm, (2) 405 nm and (3) 635 nm; 2 and 3 are the system to alter the laser used during the experiment; 4 is planar mirror; 5 is the sample holder; 6 are the biconvex lenses; 7 are the ¼ waveplate; 8 are the linear polarizer; 9 is the Ocean Optics USB4000 spectrophotometer; and 10 is the computer to control the spectrophotometer.

Table S1. Emission Ellipsometry parameters of the ZnPC samples diluted in DMSO solutions. The letters indicate the configuration of polarization of the incident light, where V is for vertical linearly polarized excitation, H is for horizontal linearly polarized excitation, L is for left-circularly polarized excitation, and R is for right-circularly polarized excitation.

| *V* | Parameters | | | | | |
|---|---|---|---|---|---|---|
| | S1/S0 | S2/S0 | S3/S0 | p | r | g |
| C3 | -0,023 | -0,012 | -0,051 | 0,057 | 0,015 | -0,102 |
| C4 | -0,009 | 0,000 | 0,001 | 0,010 | 0,006 | 0,008 |
| C5 | -0,019 | -0,004 | -0,01 | 0,023 | 0,011 | -0,016 |
| C6 | -0,020 | 0,000 | 0,001 | 0,020 | 0,013 | 0,002 |
| *H* | Parameters | | | | | |
| | S1/S0 | S2/S0 | S3/S0 | p | r | g |
| C3 | 0,015 | -0,002 | -0,039 | 0,042 | -0,010 | -0,079 |
| C4 | 0,016 | 0,002 | -0,007 | 0,017 | -0,011 | -0,013 |
| C5 | 0,007 | -0,002 | 0,002 | 0,007 | -0,005 | 0,004 |
| C6 | 0,008 | 0,005 | -0,008 | 0,012 | -0,005 | -0,016 |
| *E* | Parameters | | | | | |
| | S1/S0 | S2/S0 | S3/S0 | p | r | g |
| C3 | -0,008 | -0,033 | -0,061 | 0,070 | 0,005 | -0,121 |
| C4 | 0,013 | 0,005 | 0,002 | 0,014 | -0,009 | 0,004 |
| C5 | 0,011 | -0,002 | -0,004 | 0,012 | -0,007 | -0,008 |
| C6 | 0,014 | -0,007 | 0,001 | 0,016 | -0,009 | 0,002 |
| *D* | Parameters | | | | | |
| | S1/S0 | S2/S0 | S3/S0 | p | r | g |
| C3 | -0,030 | -0,042 | -0,045 | 0,069 | 0,020 | -0,090 |
| C4 | 0,048 | 0,000 | 0,002 | 0,047 | -0,030 | 0,009 |
| C5 | 0,016 | -0,003 | 0,000 | 0,017 | -0,011 | 0,000 |
| C6 | 0,019 | -0,005 | -0,004 | 0,020 | -0,013 | -0,008 |

*Contact author: brunozanatta@ufu.br

†Contact author: marletta@ufu.br

## III. Confocal Raman Optical Activity, experimental setup and equations.

The Raman Optical Activity (ROA) signal is obtained from the following equation [14],

$$\Delta = \frac{I_R - I_L}{I_R + I_L}, \qquad (9)$$

with $I_R$ the intensity of scattered right-hand circular polarized light, and $I_L$ the intensity of scattered left-hand circular polarized light. The scheme of the developed experimental setup is presented in Fig. S5.

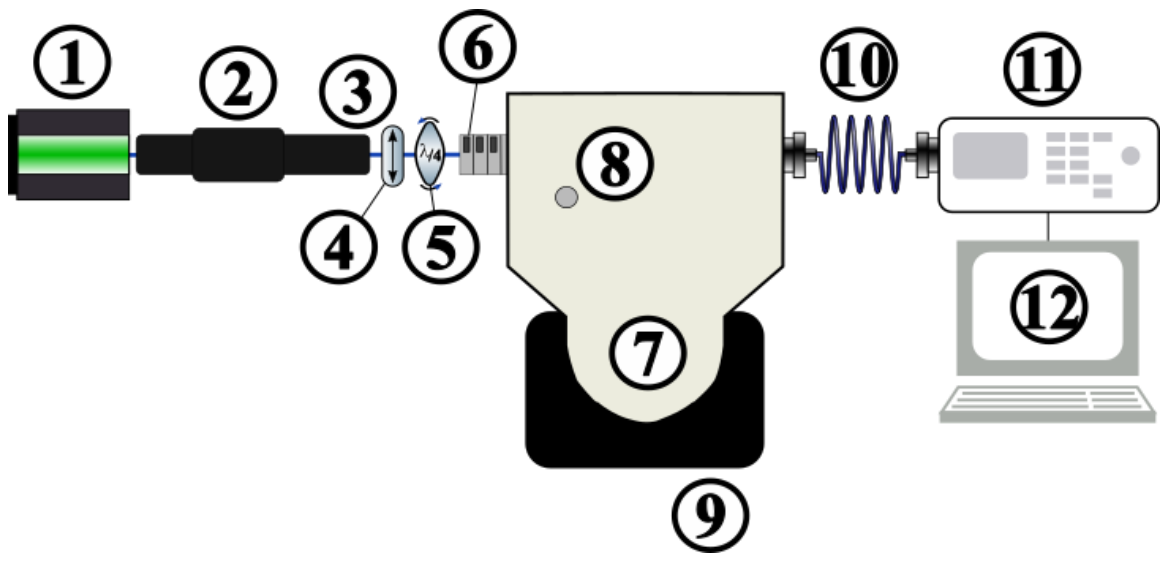

FIG. S5. Scheme to represent the Confocal-ROA setup. The numbers help identify the equipment and components used in the setup. 1 is the laser source of 514 nm; 2 is the interferometer filter for 514 nm; 3 is the PVC tube used to cover the laser light; 4 is the linear polarizer; 5 is the ¼ waveplate; 6 is the manual switch to block the laser path; 7 is the Horiba Jobin Yvon Olympus BX41 Raman microscope; 8 is the semi-lenses switch to view the live image of the sample; 9 is the table where the sample is positioned; 10 is an optical fiber; 11 is the Horiba Jobin Yvon iHR550 monochromator coupled with Synapse CCD detector; and 12 is the computer to operate the system.

## V. Line Shape Luminescence Spectrum software analysis

Theoretical modeling of the photoluminescence (PL) lineshapes was executed via the *Luminescence Spectrum Line Shape* (LSLS) framework. This algorithm evaluates vibronic transition rates under the purview of Fermi's Golden Rule, invoking both the adiabatic and Condon approximations. For a radiative decay spanning from an initial excited state *b* to the ground state *a* (*b*→*a*), the corresponding transition probability is formulated as follows [15]:

$$I(\omega) = \frac{2a_m \pi \omega^3}{3ch} |\overrightarrow{\mu_n}|^2 \times \int_{-\infty}^{+\infty} dt e^{\left(it(\omega_{ab}-\omega) - \frac{d^2t^2}{2}\right)} \prod_j G_j^*(t), \quad (10)$$

here, $\vec{\mu}_{ab}$ denotes the electronic transition dipole moment, $a_m$ accounts for local dielectric medium effects, and $c$ represents the speed of light in vacuum. The parameter $\hbar\omega_{ab} = E_b - E_a$ defines the electronic energy separation between the highest occupied molecular orbital (HOMO) and lowest unoccupied molecular orbital (LUMO) states, whereas $d$ captures the inhomogeneous spectral broadening governed by a Gaussian distribution. The function $G_j^*(t)$ corresponds to the Franck–Condon factor evaluated within the harmonic approximation, which under pure structural displacement simplifies to [15]:

$$\prod_j G_j^*(t) = exp(-\sum_{j=1}^{N} S_j[(\tilde{n}_j + 1)e^{it\omega_j} + \tilde{n}_j e^{-it\omega_j} - (2\tilde{n}_j + 1)]), \quad (11)$$

$$\tilde{n}_j = \left(e^{\frac{\hbar\omega_j}{K_B T}} - 1\right)^{-1}, \quad (12)$$

where $S_j$ denotes the Huang-Rhys factor, $\omega_j$ represents the energy of the $j^{th}$-phonon, and $\tilde{n}_j$ signifies the vibrational quantum number, with $K_B$ and $T$ defining the Boltzmann constant and sample temperature, respectively. Simulation parameters for the photoluminescence profiles were established via correlation with the experimental spectra under the strict physical constraints of the Franck–Condon model. The electronic transition energy ($E_{00} = \hbar\omega_{ab}$) was determined from the position of the emission maximum, whereas the damping/linewidth parameter ($d$) was adjusted to reproduce the observed spectral broadening arising from inhomogeneous disorder. To maintain consistency with the experimental conditions, the temperature was held constant at 300 K. Finally, the Huang–Rhys factors ($S_j$) were treated as adjustable fitting parameters, bounded within physically plausible regimes, to quantify the strength of the electron–phonon coupling.

Table S2. Adjusted parameters from the LSLS simulation of the PL spectra for samples at varying concentrations, assuming a single emitting species.

| ***Sample*** | **C6** | **C5** | **C4** | **C3** |
|---|---|---|---|---|
| **$E_{00}$ (nm)** | 680.00 | 683.00 | 689.00 | 690.00 |
| **d ($cm^{-1}$)** | 250.00 | 250.00 | 225.00 | 225.00 |
| **T (K)** | 300 | 300 | 300 | 300 |
| **$E_{01}$ ($cm^{-1}$)** | 676.00 | 676.00 | 588.00 | 588.00 |
| **$S_1$** | 0.10 | 0.10 | 0.10 | 0.07 |
| **$E_{02}$ ($cm^{-1}$)** | 1338.00 | 1200.00 | 1150.00 | 1050.00 |
| **$S_2$** | 0.05 | 0.05 | 0.10 | 0.12 |
| **$E_{03}$ ($cm^{-1}$)** | 1505.00 | 1495.00 | 1485.00 | 1480.00 |
| **$S_3$** | 0.03 | 0.03 | 0.03 | 0.03 |
| **Reliability (%)** | 96.00 | 98.00 | 97.00 | 96.00 |

*Contact author: brunozanatta@ufu.br
†Contact author: marletta@ufu.br

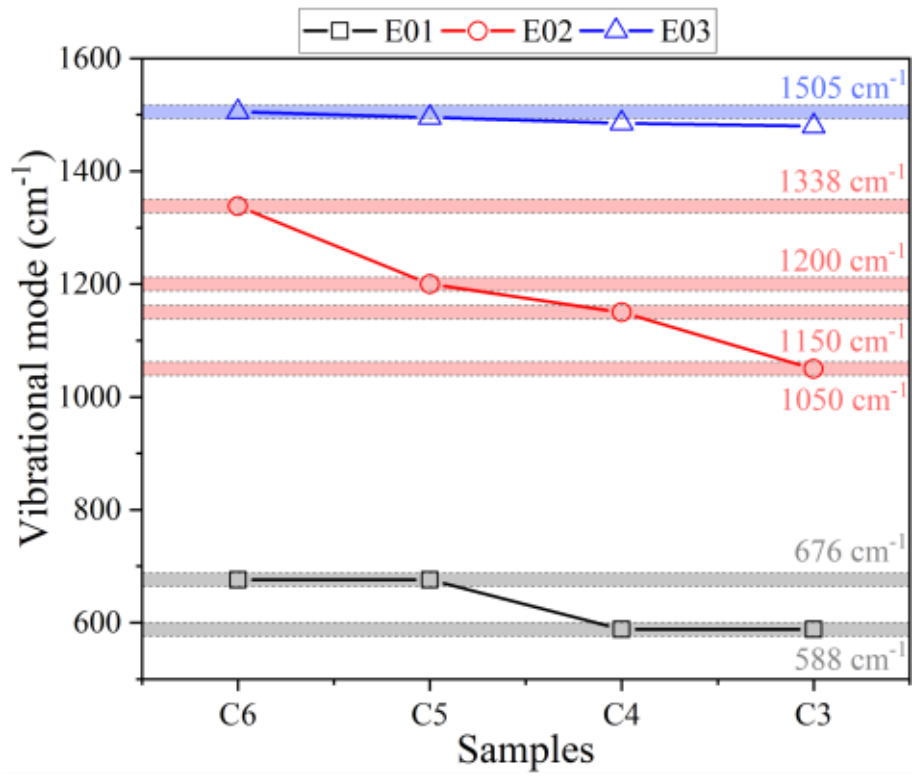

FIG. S6. Structural-vibrational mapping from Table S2, highlighting the vibronic mode reorganization and the redshift of the 1505 cm⁻¹ aza-bridge vibration.

Table S3. Adjusted parameters from the LSLS simulation assuming two emitting species.

| *Sample* | C6 | C5 | C4 | C3 |
|---|---|---|---|---|
| **Species 1** | | | | |
| **$E_{00}$ (nm)** | 680.00 | 680.00 | 680.00 | 680.00 |
| **d ($cm^{-1}$)** | 250.00 | 250.00 | 250.00 | 250.00 |
| **T (K)** | 300 | 300 | 300 | 300 |
| **$E_{01}$ ($cm^{-1}$)** | 676.00 | 676.00 | 676.00 | 676.00 |
| **$S_1$** | 0.10 | 0.10 | 0.10 | 0.10 |
| **$E_{02}$ ($cm^{-1}$)** | 1338.00 | 1338.0 0 | 1338.00 | 1338.0 0 |
| **$S_2$** | 0.05 | 0.05 | 0.05 | 0.05 |
| **$E_{03}$ ($cm^{-1}$)** | 1505.00 | 1505.0 0 | 1505.00 | 1505.0 0 |
| **$S_3$** | 0.03 | 0.03 | 0.03 | 0.03 |
| **$K_1$** | 1.00 | 1.00 | 0.25 | 0.00 |
| **Species 2** | | | | |
| **$E_{00}$ (nm)** | 690.00 | 690.00 | 690.00 | 690.00 |
| **d ($cm^{-1}$)** | 225.00 | 225.00 | 225.00 | 225.00 |
| **T (K)** | 300 | 300 | 300 | 300 |
| **$E_{01}$ ($cm^{-1}$)** | 588.00 | 588.00 | 588.00 | 588.00 |
| **$S_1$** | 0.07 | 0.07 | 0.07 | 0.07 |
| **$E_{02}$ ($cm^{-1}$)** | 1050.00 | 1050.0 0 | 1050.00 | 1050.0 0 |
| **$S_2$** | 0.12 | 0.12 | 0.12 | 0.12 |
| **$E_{03}$ ($cm^{-1}$)** | 1480.00 | 1480.0 0 | 1480.00 | 1480.0 0 |
| **$S_3$** | 0.03 | 0.03 | 0.03 | 0.03 |
| **$K_2$** | 0.00 | 0.50 | 1.00 | 1.00 |
| **Reliability (%)** | 96.00 | 90.00 | 91.00 | 96.00 |

## VI. Raman vibrational modes

Table S4. Literature-assigned Raman vibrational modes for the Zinc Phthalocyanine molecule [8-9,19-20].

| ***Frequency ($cm^{-1}$)*** | **Description of vibrational mode** |
|---|---|
| **588** | In-plane deformation of the benzene rings. |
| **676** | Macrocycle breathing; coordinated radial expansion and contraction of the entire inner ring. |
| **1050** | In-plane bending of the C–H bonds coupled with isoindole ring deformation. |
| **1150** | In-plane bending of the C–H bonds located on the peripheral benzene rings. |
| **1200** | Isoindole ring breathing and in-plane stretching of the C–C and C–H bonds of the macrocycle. |
| **1338** | Symmetric stretching of the pyrrole ring (involving internal C–N and C–C bonds). |
| **1505** | Stretching of the aza-bridges (C–N–C bonds) connecting the four phthalocyanine sub-rings. |

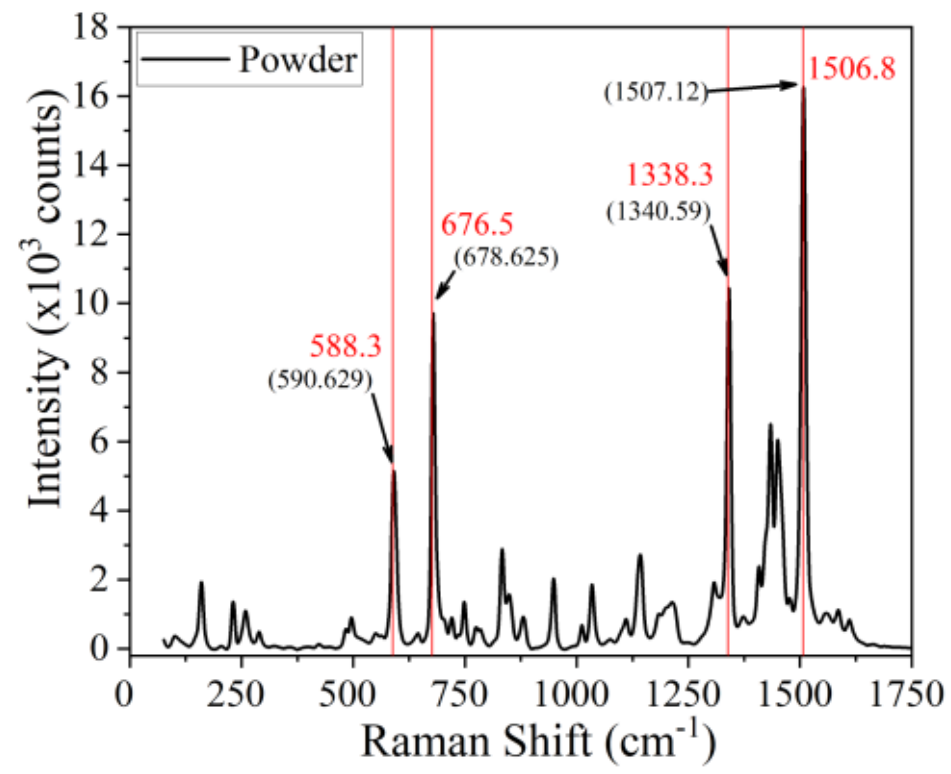

FIG. S7. Assignment of Raman-active vibrational modes for zinc phthalocyanine (ZnPc) benchmarked against literature values [20].

## VI. Density Functional Theory data

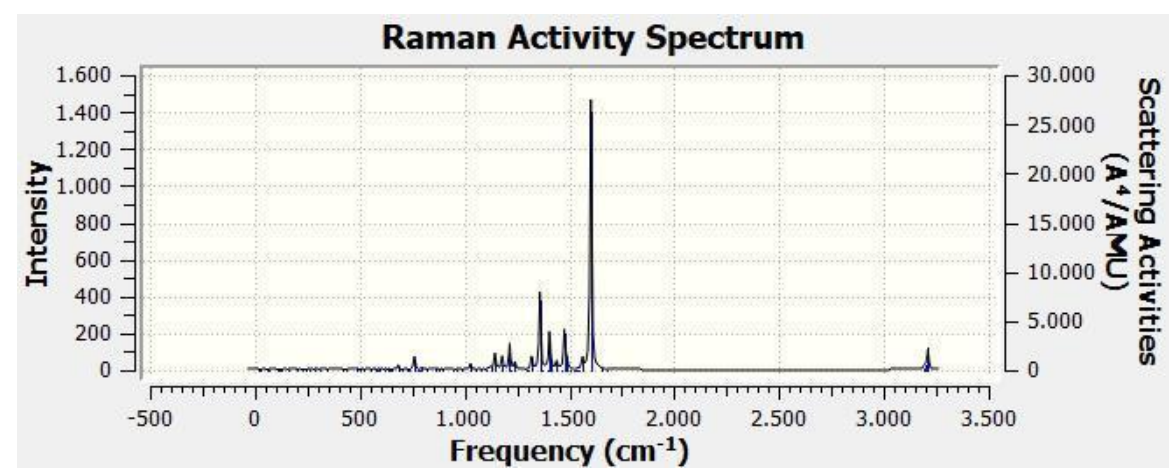

FIG. S8. Raman activity spectrum obtained via DFT simulations using the M06 functional in conjunction with the DGDZVP basis set, considering the isolated state of the ZnPc.

*Contact author: brunozanatta@ufu.br

†Contact author: marletta@ufu.br

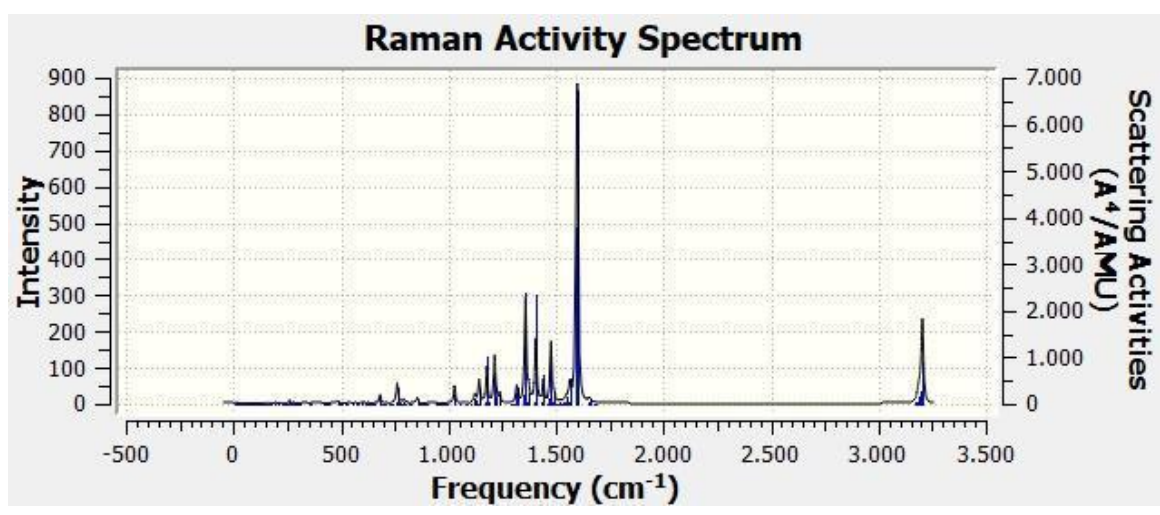

FIG. S9. Raman activity spectrum obtained via DFT simulations using the M06 functional in conjunction with the DGDZVP basis set, considering the aggregate state of the ZnPc.

Table S5. Geometrical parameters obtained for ZnPC. Vibrational frequencies and Raman intensities were also calculated for the tetramer using the energy minimized structure, the same functional and basis set.

| | Experimental* | Theoretical (M06)* | Theoretical $D_{4h}$ (B3LYP/6-311++G(2d,2p)) * |
|---|---|---|---|
| $C_2$- $C_{2a}$ | 1.398 | 1.404 | 1.407 |
| Zn-$N_1$ | 1.980 | 1.990 | 1.999 |
| $N_1$-$C_1$ | 1.366 | 1.367 | 1.369 |
| $C_1$-$N_2$ | 1.332 | 1.327 | 1.327 |
| $C_1$-$C_2$ | 1.450 | 1.455 | 1.459 |
| $C_2$-$C_3$ | 1.393 | 1.392 | 1.391 |
| $C_3$-$C_4$ | 1.388 | 1.391 | 1.389 |
| $C_4$-$C_{4a}$ | 1.398 | 1.405 | 1.403 |
| Zn- $N_1$-$C_1$ | 124.6 | 125.1 | 125.0 |
| $N_1$-$C_1$-$N_2$ | 127.5 | 127.8 | 127.5 |
| $C_1$-$N_1$-$C_{1a}$ | 109.4 | 109.7 | 109.9 |
| $N_1$-$C_1$-$C_2$ | 108.4 | 108.5 | 108.4 |
| $C_1$-$C_2$-$C_{2a}$ | 106.3 | 106.6 | 106.6 |
| $C_2$-$C_{2a}$-$C_{3a}$ | 120.7 | 121.2 | 121.0 |
| $C_2$-$C_3$-$C_4$ | 117.6 | 117.6 | 117.8 |
| $C_3$-$C_4$-$C_{4a}$ | 121.5 | 121.2 | 121.1 |

*Bond lengths in Å, angles in degrees.

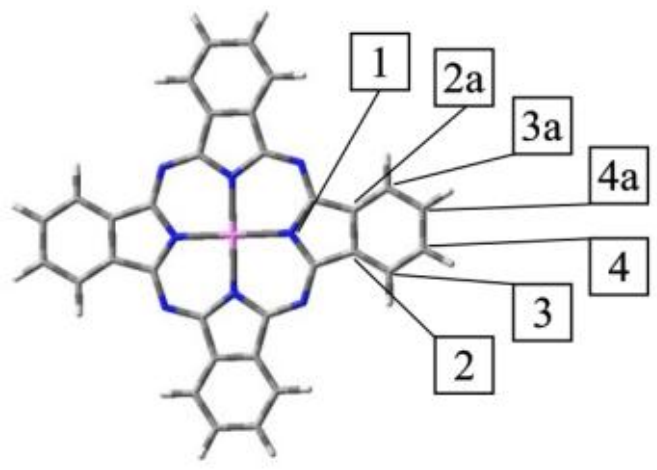

FIG. S10. Representation of the position of the atoms in the molecule, as seen in Table S5.

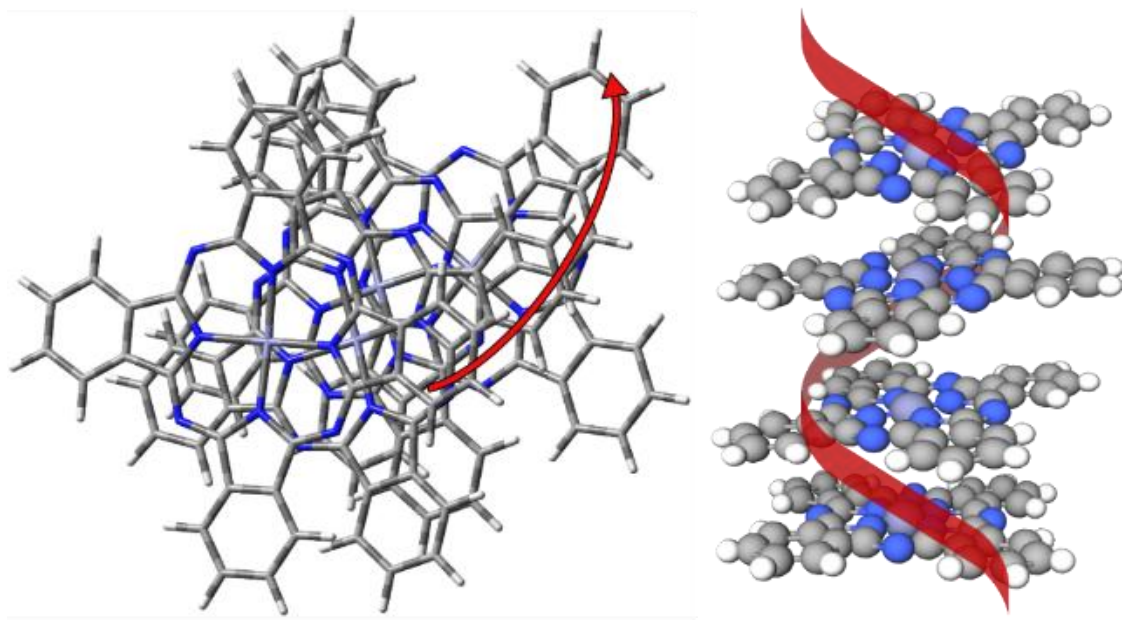

FIG S11. Three-dimensional representation of the ZnPc tetrameric aggregate derived from the DFT simulation (left), with the red arrow indicating the symmetry-breaking profile that forms a left-handed helix. A simplified helical representation (right) aids in visualizing the emergent chiral architecture.

*Contact author: brunozanatta@ufu.br
†Contact author: marletta@ufu.br